\documentclass[aps,prc,letterpaper,11pt
,twoside,tightenlines,nofootinbib,showpacs
,preprint]{revtex4-1}
\usepackage{graphicx}
\usepackage{color}
\usepackage[sort&compress]{natbib}
\usepackage{subfigure}
\usepackage{amsmath}
\usepackage{amssymb}
\usepackage{amsfonts}
\usepackage{cancel}
\usepackage{hyperref}
\usepackage{caption}
\usepackage{fontenc}

\begin{document}

\arraycolsep1.5pt

\newcommand{\Ima}{\textrm{Im}}
\newcommand{\Rea}{\textrm{Re}}
\newcommand{\mev}{\textrm{ MeV}}
\newcommand{\be}{\begin{equation}}
\newcommand{\ee}{\end{equation}}
\newcommand{\ba}{\begin{eqnarray}}
\newcommand{\ea}{\end{eqnarray}}
\newcommand{\gev}{\textrm{ GeV}}
\newcommand{\nn}{{\nonumber}}
\newcommand{\dtres}{d^{\hspace{0.1mm} 3}\hspace{-0.5mm}}
\raggedbottom

\def\del{\partial}

\title{The composite nature of the $\Lambda(1520)$ resonance}

\author{F. Aceti$^1$, E. Oset$^1$ and L. Roca$^2$}
\affiliation{
$^1$ Departamento de F\'{\i}sica Te\'orica and IFIC, Centro Mixto 
Universidad de Valencia-CSIC,
Institutos de Investigaci\'on de Paterna, Aptdo. 22085, 46071 Valencia,
Spain\\
$^2$ Departamento de F\'\i sica, Universidad de Murcia, E-30100 Murcia,
Spain.
 }

\date{\today}

\begin{abstract}

Recently, the Weinberg  compositeness condition of a bound state was
generalized to  account for resonant states and higher partial waves. We
apply this extension to the case of the $\Lambda(1520)$ resonance and
quantify the weight of the  meson-baryon components in contrast to
other possible genuine building blocks. This resonance was 
theoretically  obtained from a coupled channels analysis using  the
s-waves $\pi\Sigma^*$, $K\Xi^*$ and the d-waves $\bar{K}N$ and
$\pi\Sigma$ channels applying the techniques of the chiral unitary
approach. We obtain that this resonance is essentially dynamically
generated from these meson-baryon channels, leaving room for only $15 \%$
weight of other kind of components into its wave function.

\end{abstract}

\maketitle


\section{Introduction}
\label{Intro}

One of the most important issues in hadron spectroscopy is the
determination of the nature of different hadronic states, mesons and
baryons, found in different facilities and reported in the PDG
\cite{pdg}, and the description of the spectrum of excited hadrons.

It has become clear that the traditional idea that considers mesons and
baryons as pure $q\bar{q}$ and $qqq$ quark states, respectively, has to be
replaced in some cases by more complex pictures involving more than two
or three quarks. In the last years, the application of Chiral
Perturbation Theory ($\chi PT$) to the study of the interactions of
hadrons \cite{chiralwein,chiralgasser} had remarkable success in
describing hadron structures. However, this effective field theory, in
which the ground states of mesons and baryons are considered as 
the relevant degrees
of freedom, is not suitable to deal with the problem of spectroscopy,
due to its very limited range of convergence.

The method has been improved constructing a non-perturbative unitary extension of the theory, called chiral unitary approach \cite{Kaiser:1995eg, Kaiser:1996js, npa, Kaiser:1998fi, ramonet, angelskaon, ollerulf, Jido:2002yz, carmen, cola, carmenjuan, hyodo, review}, which allows to explain many mesons and baryons as composite states of hadrons. This kind of resonances are commonly known as ``dynamically generated".

One of the most challenging issues in this field is to understand whether a resonance can be considered as composite of other hadrons or something different, eventually a genuine resonance. The first breakthrough in the investigation of the composite nature of a system of particles was made in 1965 by Weinberg in the well-known paper in which he determined that the deuteron was a bound state of a proton and a neutron \cite{weinberg}. The same issue was studied later in \cite{han1, han2, cleven}. However, the method was only suitable for $s$-waves and small binding energies.

A generalization to more heavily bound systems and using many coupled
channels was made  in \cite{gamermann} and extended to the case of
resonances in \cite{yamagata}, while in \cite{aceti} the analysis was 
modified to include any partial waves. The method contained in
\cite{aceti} was successfully applied to the $\rho$ meson in that work,
confirming the commonly accepted idea that it is not a $\pi\pi$
composite state but a genuine resonance, and in \cite{xiao} to the $K^*$
meson, finding a very small weight for the $K\pi$ component in its wave
function.

This generalization to any partial waves and resonant states of the Weinberg compositeness condition found in \cite{aceti}, was applied to baryons in \cite{aceti2} for the first time in order to determine the weight of the meson-baryon component in the members of the $J^P=\frac{3}{2}^+$ baryons decuplet. An amount of $60\%$ has been found for the $\pi N$ cloud in the $\Delta(1232)$, while the higher-energy members of the decuplet seem to be better represented by a genuine component.

In the present work we investigate the structure of another baryonic
resonance, the $\Lambda(1520)$. This resonance belongs to the negative
parity $J^P=\frac{3}{2}^-$ resonances that, in the last few years, have
been interpreted as dynamically generated from the interaction of the
octet of the pseudoscalar mesons with the decuplet of the baryons
\cite{lambda1, lambda2}. It has been studied theoretically in
\cite{lambda1, lambda2} and  considered as generated from the
interaction of the coupled channels $\pi\Sigma(1385)$ and $K\Xi(1530)$
in s-wave.
In this picture, it couples mostly to the first channel, qualifying as a
quasi-bound state of $\pi\Sigma^*$, with a nominal mass  of few MeV
below the $\pi\Sigma^*$ threshold. However, the large branching ratios 
to $\bar{K}N$ and $\pi\Sigma$ indicate that these two channels must play
a remarkable role in the building up of the resonance in spite of the
fact that they couple in d-wave.

In \cite{luislambda1} a coupled channels analysis of the $\Lambda(1520)$
data using $\pi\Sigma^*$, $K\Xi^*$, $\bar{K}N$ and $\pi\Sigma$ has been
performed. In this work, the $\pi\Sigma^*$ channel is still the one with
the largest coupling, but its strength is reduced with respect to the
quasibound $\pi\Sigma^*$ picture. At the same time, the couplings to
$\bar{K}N$ and $\pi\Sigma$ are remarkable, making these two channels
relevant for the interpretation of different reactions involving the
$\Lambda(1520)$. The model provided in \cite{luislambda1} has been
tested in \cite{luislambda2} through the study of the two reactions
$pp\rightarrow pK^+K^-p$ and $pp\rightarrow pK^+\pi^0\pi^0\Lambda$ close
to the $\Lambda(1520)$ threshold, giving important information about the
couplings of the $\Lambda(1520)$ to $\bar{K}N$ and $\pi\Sigma^*$.

In this work, by means of the extension of the Weinberg sum rule to resonant states in any partial waves, we make an estimation of the relevance of the different channels in the wave function of the $\Lambda(1520)$, starting from the coupled channel study of Ref. \cite{luislambda1} and \cite{luislambda2}.


\section{Summary of the formalism and meaning of the sum rule for resonances}
\label{form}

For the sake of completeness, let us first briefly summarize the
approach used  in Ref.~\cite{aceti} to study the composite nature of a
resonance in any partial waves. In order to create a resonance from the
interaction of many channels at a certain energy, two particles must
collide in a channel which is open at this energy. The process is
described by the set of coupled Schr\"{o}dinger equations,

\begin{equation}
\label{eq:ls_res}
\begin{split}
|\Psi\rangle&=|\Phi\rangle +\frac{1}{E-H_{0}}V|\Psi\rangle\\&=|\Phi\rangle +\frac{1}{E-M_i-\frac{\vec{p}\,^2}{2\mu_i}}V|\Psi\rangle\ ,
\end{split}
\end{equation}
where
\begin{equation}
\label{eq:2}
|\Psi\rangle=
\begin{Bmatrix}
|\Psi_{1}\rangle \\ |\Psi_{2}\rangle \\\vdots \\|\Psi_{N}\rangle
\end{Bmatrix}\ ,\ \ \ \ \ \ \
|\Phi\rangle=
\begin{Bmatrix}
|\Phi_{1}\rangle \\0\\\vdots \\0
\end{Bmatrix}\ ,
\end{equation}
$H_0$ is the free Hamiltonian and $\mu_{i}$ is the reduced mass of the system of total mass $M_{i}=m_{1i}+m_{2i}$. The state $|\Phi_1\rangle$ is an asymptotic scattering state used to create a resonance which will decay into other channels.

The potential $V$ has the form
\begin{equation}
\label{eq:pot}
\langle\vec{p}\,|V|\vec{p}\ '\rangle\equiv (2l+1)\ v\ \Theta(\Lambda-p)\Theta(\Lambda-p')|\vec{p}\,|^{l}|\vec{p}\ '|^{l}P_{l}(\cos\theta)\ ,
\end{equation}
where $\Lambda$ is a cutoff in the momentum space and $v$ is a $N\times N$ matrix, with $N$ the number of channels. The form of the potential is such that the generic $l$-wave character of the process is factorized in $|\vec{p}\,|^{l}$ and $|\vec{p}\ '|^{l}$, and in the Legendre polynomial $P_{l}(\cos\theta)$. Thus, the matrix $v$ is a constant matrix.

The $N\times N$ scattering matrix such that $T\Phi=V\Psi$, can be written as
\begin{equation}
\label{eq:tt}
T=(2l+1)P_{l}(\hat{p},\hat{p}')\Theta(\Lambda-p)\Theta(\Lambda-p')|\vec{p}\,|^{l}|\vec{p}\ '|^{l}t\ ,
\end{equation}
and the Schr\"odinger equation leads to the Lippmann-Schwinger equation for $T$ ($T=V+VGT$), by means of which one obtains
\begin{equation}
\label{eq:t}
t=\frac{v}{(1-vG)}=\frac{1}{v^{-1}-G}\ . 
\end{equation}
The matrix $G$ in Eq.~\eqref{eq:t} is a diagonal matrix accounting for 
the two hadron loop functions in the intermediate state (see Eq.~\eqref{eq:loop}). 
It is important to stress that there is no explicit $\Lambda(1520)$ pole
included into the formalism, and thus the
resonant shape (see Fig.~\ref{fig:fit}) and pole comes from the non-linear
dynamics involved in the unitarization.

On the other hand, note that the
definition  $T\Phi=V\Psi$ makes $T$ independent of the phase convention of the wave function.
This derivation leads to a $t$ matrix which does not contain the factor $|\vec{p}\,|^{l}$, due to the constant $v$ matrix. Other approaches for $p$-waves, like the ones in \cite{ollerpalomar,doring}, factorize on shell $|\vec{p}\,|^{l}$ and associate it to the potential $v$. In this new approach, this factor is absorbed in a new loop function
\begin{equation}
\label{eq:loop}
G_{ii}=\int_{_{|\vec{p}\,|<\Lambda}}{d^3p\,\frac{|\vec{p}\ |^{2l}}{E-m_{1i}-m_{2i}-\frac{\vec{p}\,^{2}}{2\mu_i}+i\epsilon}}\ ,
\end{equation}
which is different from the one normally used in the chiral unitary approach \cite{oller}. 

This choice is necessary for the generalization of the sum rule for the couplings to any partial wave found in \cite{aceti}, which holds both for resonances and bound states dynamically generated by the interaction in coupled channels of two hadrons,
\begin{equation}
\label{eq:sumrule}
\sum_{i}g_{i}^{2}\left[\frac{dG_{i}}{dE}\right]_{E=E_{R}}=-1\ .
\end{equation}  
In Eq.~\eqref{eq:sumrule}, $E_R$ is the position of the complex pole
of the scattering matrix in the second Riemann sheet (see definition
below)
 representing the resonance and $g_i$ is the coupling to the channel $i$ defined as
\begin{equation}
\label{eq:coupling}
g_ig_j=\lim_{E\rightarrow E_R}(E-E_R)t_{ij}\ .
\end{equation}

Note that this definition leads to complex couplings. This means that the terms of Eq.~\eqref{eq:sumrule} are complex, which implies that the imaginary parts cancel and one is left with 
\begin{equation}
\label{eq:srreal}
\sum_{i}Re\left(g_{i}^{2}\left[\frac{dG_{i}}{dE}\right]_{E=E_{R}}\right)=-1\ .
\end{equation}

Each term in Eq.~\eqref{eq:sumrule} represents the integral of the wave function squared (not the modulus squared as it was in the case of bound states \cite{gamermann}) of each component,
\begin{equation}
\label{eq:wf2}
\int d^3p\,(\Psi_i(p))^2=-g_i^2\frac{\partial G_i^{II}}{\partial E}\ ,
\end{equation}
where the superindex $II$ stands for second Riemann
sheet,
 but this occurs only with the phase convention in which the wave function in momentum space is given by
\begin{equation}
\label{eq:wf}
\Psi_i(p)=g_i\frac{\Theta(\Lambda-|\vec{p}\,|)p^{\,l}}{E-m_{1i}-m_{2i}-p^2/2\mu_i+i\epsilon}\ .
\end{equation}
This is a most appropriate choice, in which the radial wave function is
real for a bound state.

We can consider each one of the terms in Eq.~\eqref{eq:srreal} as a measure of the relevance, or the weight, of a channel in the wave function of the state, but not a probability, which for open channels is not a useful concept since it will diverge, as explained in detail in \cite{aceti2}.

Sometimes, we only have information on hadron-hadron scattering. There can be a genuine component  different to the hadron-hadron one and, in order to take into account its weight,
 Eq.~\eqref{eq:srreal} 
must be rewritten as
\begin{equation}
\label{eq:sumrule2}
-\sum_{i}Re\left(g_{i}^{2}\left[\frac{dG_{i}}{dE}\right]_{E=E_{R}}\right)=1-Z\ ,\ \ \ \ \ \ \ \ \ \ \ \ Z=Re\int d^3p\left(\Psi_{\beta}(p)\right)^2\ ,
\end{equation}
where $\Psi_{\beta}(p)$ is the genuine component in the wave function of the state, when it is omitted from the coupled channels \cite{hyodorep}.

The left-hand side of Eq.~\eqref{eq:sumrule2} is the measure of the relevance of the hadron-hadron component, while its diversion from unity, $Z$, measures the weight of something different in the wave function.


\section{Application to the $\mathbf{\Lambda(1520)}$ resonance}

As mentioned in the introduction,  the method of Sec.~\ref{form} has
already been applied to the $\rho$ and $K^*$ mesons in \cite{aceti} and
\cite{xiao} respectively, and, for the first time, the structure of
baryonic resonances has been investigated in \cite{aceti2}, making an
estimation of the meson-baryon component in the wave function of the
baryons of the decuplet of the $\Delta(1232)$. 

We apply now the method to another baryonic resonance, the $\Lambda(1520)$, with quantum numbers $J^P=\frac{3}{2}^-$.


\subsection{The chiral unitary model for the $\mathbf{\Lambda(1520)}$}
\label{model}
Our starting point is the analysis of Ref. \cite{luislambda1}. The $\Lambda(1520)$ is studied in the framework of a coupled channels formalism including the channels $\pi\Sigma(1385)$ and $K\Xi(1530)$ in $s$-waves and $\bar{K}N$ and $\pi\Sigma$ in $d$-waves.

The matrix containing the tree-level amplitudes can be written as \begin{equation} V=\left(
\begin{array}{cccc} \mathcal{C}_{11}(k_1^0+k_1^0)\ \ \  & \ \ \ \mathcal{C}_{12}(k_1^0+k_2^0)\ \ \ 
& \gamma_{13}\,q_3^2 & \ \ \ \gamma_{14}\,q_4^2  \\ \mathcal{C}_{21}(k_1^0+k_2^0)\ \ \  &\ \ \
\mathcal{C}_{22}(k_2^0+k_2^0)\ \ \  & 0 & \ \ \ 0  \\ \gamma_{13}\,q_3^2 & 0 & \gamma_{33}\,q_3^4 & 
\ \ \ \gamma_{34}\,q_3^2\,q_4^2  \\ \gamma_{14}\,q_4^2 & 0 & \gamma_{34}\,q_3^2\,q_4^2 & \ \ \
\gamma_{44}\,q_4^4  \end{array} \right)\ , \label{eq:potential_luis} \end{equation}
where $q_i=\sqrt{(s-(m_i-M_i)^2)(s-(m_i+M_i)^2)}/2\sqrt{s}$ and $k_i^0=(s-M_i^2+m_i^2)/2\sqrt{s}$,
 with $m_i$ and $M_i$ the masses of the meson and baryon in channel $i$ ($i=1,4$), respectively. The
 $s$- and $d$-waves character of the transitions is taken into account by means of the dependence of
 the potentials on the incoming and outgoing squared momenta. 
The s-wave transition elements are 
obtained
 from the lowest order chiral Lagrangian involving the interaction of the
decuplet of baryons and the octet of pseudoscalar mesons, which gives
$\mathcal{C}_{11}=-1/f^2$, $\mathcal{C}_{12}=\mathcal{C}_{21}=-\sqrt{6}/4f^2$ and
 $\mathcal{C}_{22}=-3/4f^2$, with  $f=1.15\ f_{\pi}$ and $f_{\pi}=93$ MeV. The factor $1.15$ in $f$
 represents the average between $f_{\pi}$ and $f_K$ \cite{angelskaon}.

The scattering amplitudes are then evaluated by means of the Bethe-Salpeter equation, which reads
\begin{equation}
T=[1-V\,G]^{-1}\ V\ .
\label{eq:BS}
\end{equation}

As explained in detail in \cite{aceti}, the method summarized
in Sec.~\ref{form} requires a potential independent of the momenta
of the particles, which is not the case of Eq.~\eqref{eq:BS}. 
For this reason we define a new potential

\begin{equation}
V'=\left(
\begin{array}{cccc}
\mathcal{C}_{11}(k_1^0+k_1^0)\ \ \  & \ \ \ \mathcal{C}_{12}(k_1^0+k_2^0)\ \ \  & \gamma_{13} & \ \ \ \gamma_{14}  \\
\mathcal{C}_{21}(k_1^0+k_2^0)\ \ \  &\ \ \ \mathcal{C}_{22}(k_2^0+k_2^0)\ \ \  & 0 & \ \ \ 0  \\
\gamma_{13} & 0 & \gamma_{33} &  \ \ \ \gamma_{34}  \\
\gamma_{14} & 0 & \gamma_{34} & \ \ \ \gamma_{44} 
\end{array}
\right)\ 
\label{eq:potential_indep}
\end{equation}
independent of the $q^2$ factors from the d-waves and, 
according to Sec. \ref{form}, we include this dependence in the new loop functions, that will now have the form
\begin{equation}
G^{(s)}_{i}=2M_i\int\frac{d^3p}{(2\pi)^3}\frac{\omega_{i}(p)+E_{i}(p)}{2\omega_{i}(p)E_{i}(p)}\frac{1}{P^{02}-(\omega_{i}(p)+E_{i}(p))^2+i\epsilon}\ ,
\label{eq:loop_swave}
\end{equation}
\begin{equation}
G^{(d)}_{i}=2M_i\int\frac{d^3p}{(2\pi)^3}\frac{\omega_{i}(p)+E_{i}(p)}{2\omega_{i}(p)E_{i}(p)}\frac{p^4}{P^{02}-(\omega_{i}(p)+E_{i}(p))^2+i\epsilon}\ ,
\label{eq:loop_dwaveold}
\end{equation}
where $\omega_i(p)=\sqrt{p^2+m_i^2}$ and $E_i(p)=\sqrt{p^2+M_i^2}$ are the energies of the meson and the baryon involved in the loop, respectively. 
This modification concerns only the loop function for the $d$-wave
channels, $G_i^{(d)}$ ($i=3,4$), which now contains the factor
$p^{2l}=p^4$. Note that we improve on Eq.~\eqref{eq:loop} by taking the
relativistic propagator~\cite{npa}.

Note that the s-wave elements of the $V'-$matrix ($V'_{11}$, $V'_{12}$,
$V'_{21}$ and $V'_{22}$) still contain an energy dependence
through the $k^0_i$ factors. However this dependence is very smooth and
it will play a role in the diversion from unity in the sum rule 
of Eq.~\eqref{eq:sumrule2}, as will be further explained in the results
section. 

The loop functions in Eqs.~\eqref{eq:loop_swave} and
\eqref{eq:loop_dwaveold} are regularized by means of two different
cutoffs in momentum space, $p^{(s)}_{max}$ and  $p^{(d)}_{max}$ for the
s- and d-waves channels respectively.  These
two cutoffs, together with the coefficients $\gamma_{13}$,
$\gamma_{14}$, $\gamma_{44}$, $\gamma_{44}$ and $\gamma_{34}$ of
Eq.~\eqref{eq:potential_indep}, constitute the set of free parameters in
the theory.

However, this procedure presents a problem due to the different dimensions of the magnitudes
 involved. The elements of $V'$ in Eq.~\eqref{eq:potential_indep} concerning the transitions
 involving the $d$-waves channels, after removing the dependence on the momenta,  will have
 different dimension with respect to the other ones. The same happens to the loop function, that now
have different dimensions in the cases of $s$- or $d$-waves.

 In order to render the dimensions homogeneous and evaluate the scattering amplitudes by means of the Bethe-Salpeter
  equation (Eq.~\eqref{eq:BS}), which is a matrix equation, we define
 \begin{equation}
\tilde{V}=\left(
\begin{array}{cccc}
\mathcal{C}_{11}(k_1^0+k_1^0)\ \ \  & \ \ \ \mathcal{C}_{12}(k_1^0+k_2^0)\ \ \  & \gamma_{13}\,q_3^2(m_{\Lambda^*}) & \ \ \ \gamma_{14}\,q_4^2(m_{\Lambda^*})  \\
\mathcal{C}_{21}(k_1^0+k_2^0)\ \ \  &\ \ \ \mathcal{C}_{22}(k_2^0+k_2^0)\ \ \  & 0 & \ \ \ 0  \\
\gamma_{13}\,q_3^2(m_{\Lambda^*}) & 0 & \gamma_{33}\,q_3^4(m_{\Lambda^*}) &  \ \ \ \gamma_{34}\,q_3^2(m_{\Lambda^*})\,q_4^2(m_{\Lambda^*})  \\
\gamma_{14}\,q_4^2(m_{\Lambda^*}) & 0 & \gamma_{34}\,q_3^2(m_{\Lambda^*})\,q_4^2(m_{\Lambda^*}) & \ \ \ \gamma_{44}\,q_4^4(m_{\Lambda^*}) 
\end{array}
\right)\ ,
\label{eq:potential}
\end{equation}
with
$q_i(m_{\Lambda^*})=\sqrt{(m_{\Lambda^*}^2-(m_i-M_i)^2)(m_{\Lambda^*}^2-(m_i+M_i)^2)}/2m_{\Lambda^*}$,
where we choose for $m_{\Lambda^*}$  the $\Lambda(1520)$ mass.
Nevertheless this specific choice is obviously 
irrelevant for the final results.
 Now all the
elements of the matrix $\tilde{V}$ have the same dimensions. Moreover,
we can now write 

\begin{equation}
\tilde{G}^{(d)}_{i}=2M_i\int\frac{d^3p}{(2\pi)^3}\frac{\omega_{i}(p)+E_{i}(p)}{2\omega_{i}(p)E_{i}(p)}\frac{p^4/q_i^4(m_{\Lambda^*})}{P^{02}-(\omega_{i}(p)+E_{i}(p))^2+i\epsilon}\ .
\label{eq:loop_dwave}
\end{equation}
This new loop function for the $d$-waves cases has the same dimension as $G_i^{(s)}$, and now the Bethe-Salpeter equation reads
\begin{equation}
\tilde{T}=\frac{1}{\tilde{V}^{-1}-\tilde{G}}\ ,
\label{eq:bethesalpeter}
\end{equation}
with 
\begin{equation}
\tilde{G}=\left(
\begin{array}{cccc}
G_1^{(s)}  & 0  & 0 & 0 \\
0 & G_2^{(s)} & 0 &  0  \\
0 & 0 & \tilde{G}_3^{(d)} &  0  \\
0 & 0 & 0 & \tilde{G}_4^{(d)} 
\end{array}
\right)\ .
\end{equation}


\section{Results}
\label{results}
The scattering amplitudes derived using Eq.~\eqref{eq:bethesalpeter} contain, as we already mentioned, seven free parameters: $p^{(s)}_{max}$, $p^{(d)}_{max}$, $\gamma_{13}$, $\gamma_{14}$, $\gamma_{44}$, $\gamma_{44}$ and $\gamma_{34}$. In order to obtain their values, we fit the model to the
 experimental scattering amplitudes for $\bar{K}N$ and $\pi\Sigma$ in $d$-wave and for $I=0$ \cite{data1, data2}. The relation between the experimental and the theoretical amplitudes is given by
\begin{equation}
T^{exp}_{ij}(\sqrt{s})=-\sqrt{\frac{M_iq_i}{4\pi\sqrt{s}}}\sqrt{\frac{M_jq_j}{4\pi\sqrt{s}}}\,T_{ij}(\sqrt{s})\ ,
\label{eq:exp-the}
\end{equation}
with $i$ and $j$ the channels involved in the transition.

\begin{figure}
  \centering
  \subfigure[]{\label{fig:fit1}\includegraphics[width=8.cm]{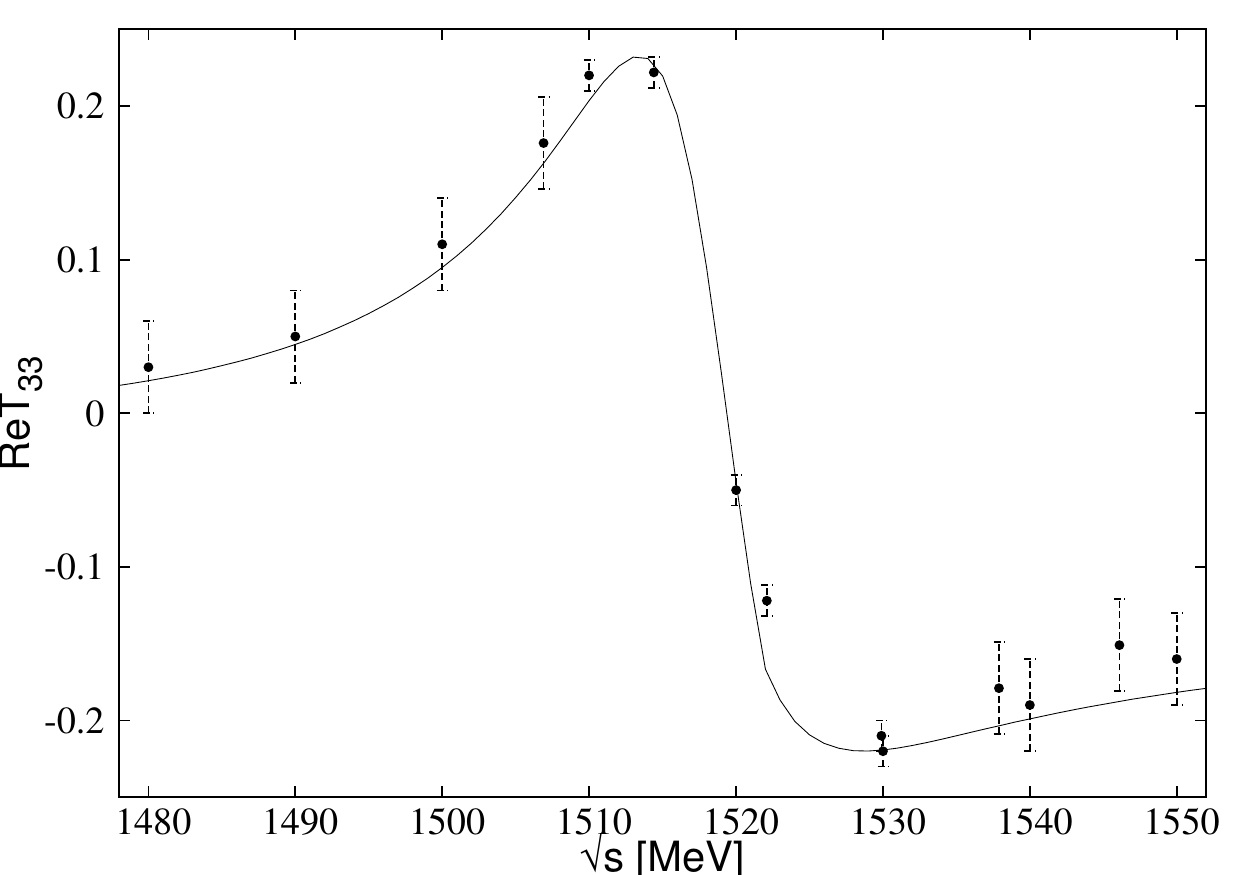}}                
   \subfigure[]{\label{fig:fit3}\includegraphics[width=8.cm]{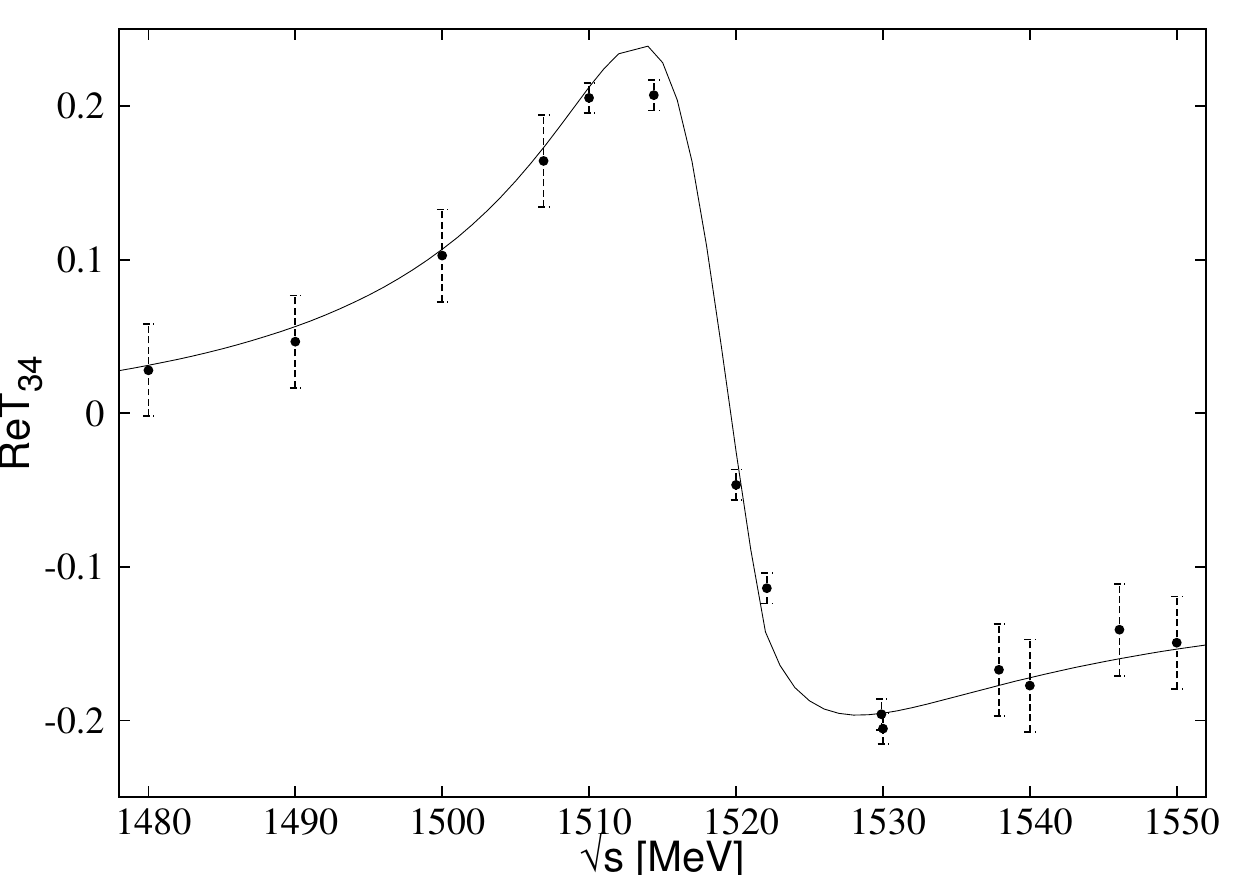}}\\
   \subfigure[]{\label{fig:fit1}\includegraphics[width=8.cm]{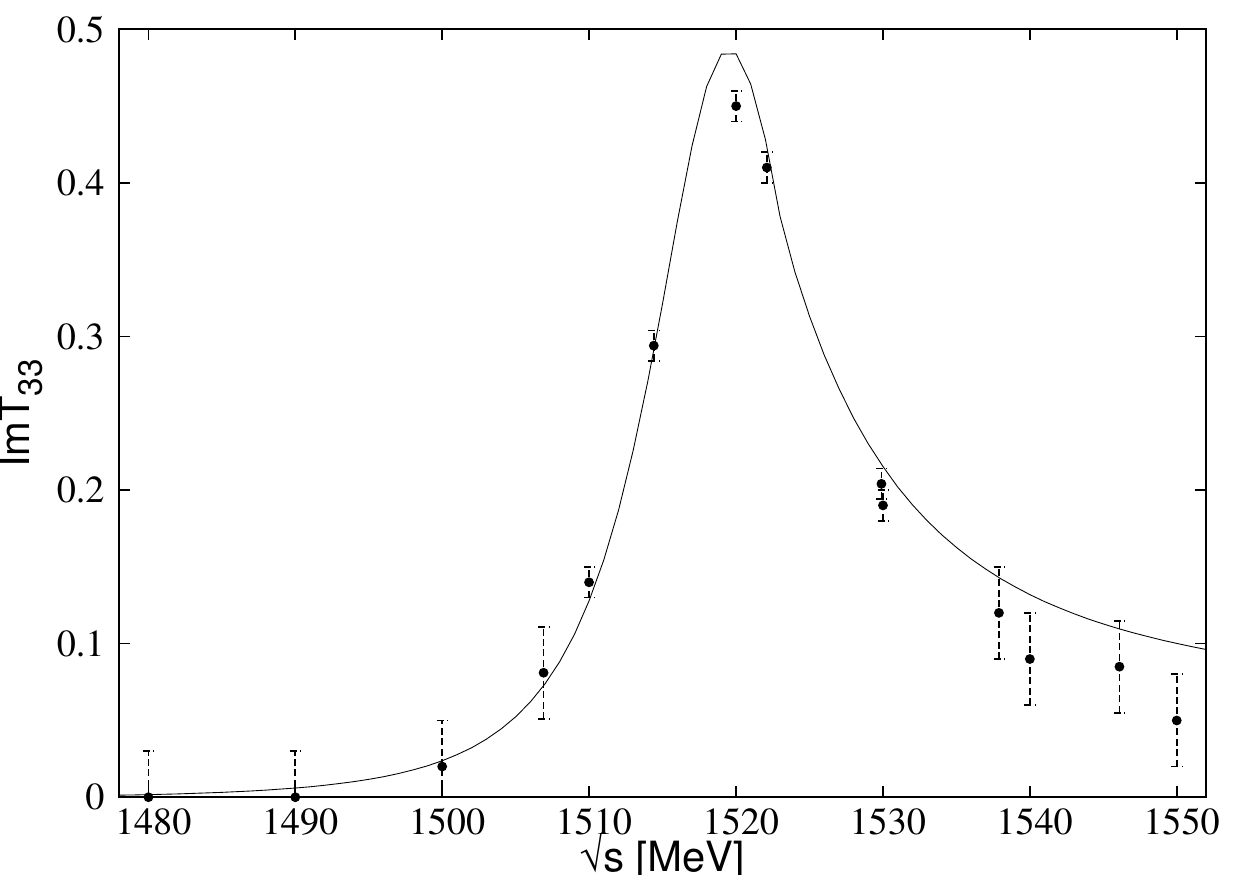}}                
   \subfigure[]{\label{fig:fit3}\includegraphics[width=8.cm]{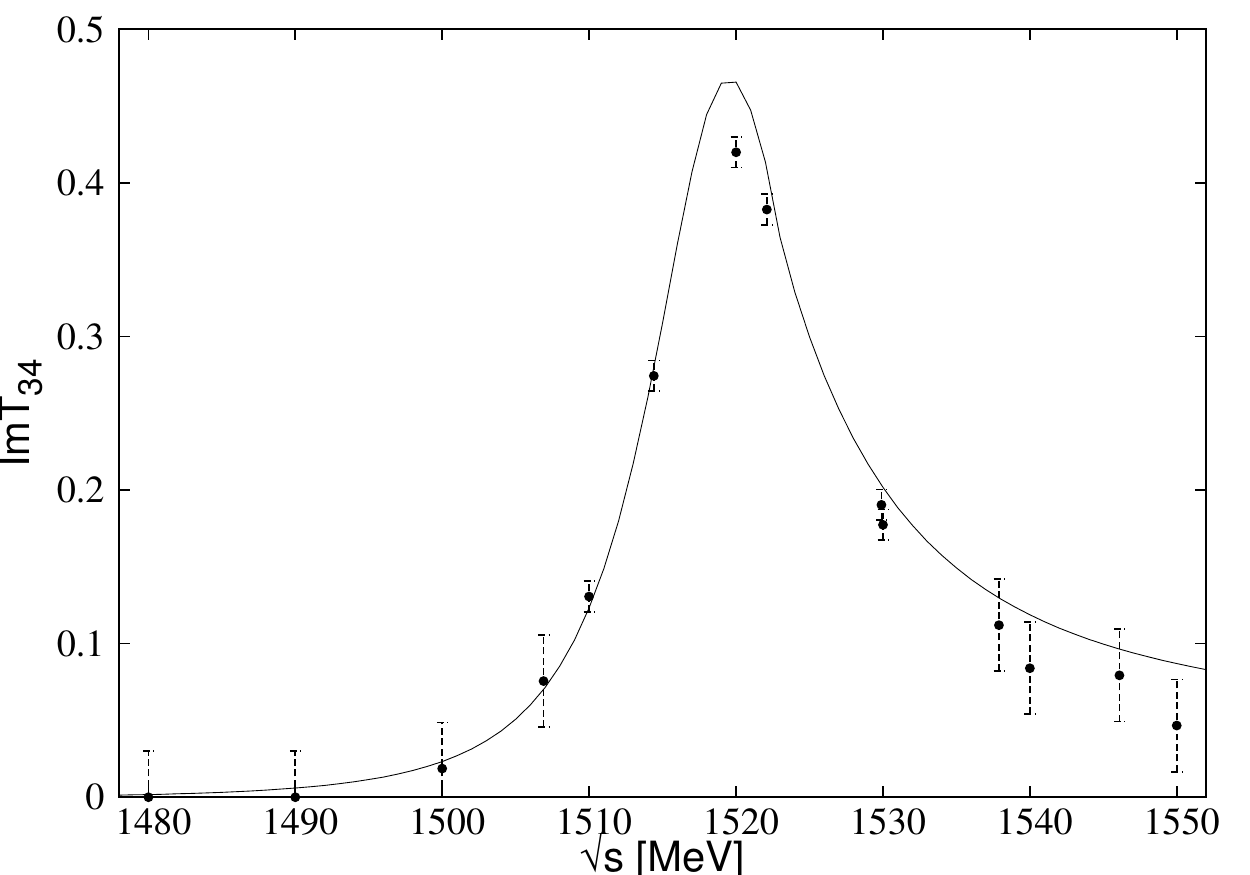}}
  \caption{Fit N. 1) to the experimental amplitudes for the transitions $\bar{K}N\rightarrow\bar{K}N$ in Figs. $a)$ and $c)$, and $\bar{K}N\rightarrow\pi\Sigma$ in Figs. $b)$ and $d)$.}
  \label{fig:fit}
\end{figure}

We obtain several equivalent best fits to the experimental data and, in
 Tab.~\ref{tab:fit} the values of the parameters obtained from a sample of
five sets are listed. In Fig.~\ref{fig:fit} we show the results of the
fit for the first set of Tab.~\ref{tab:fit}, but, in any case, the
results are consistent for all the sets. 
The fact that we get approximately the same solutions with different 
sets of parameters indicates that there are strong
correlations between them. The final results are also very similar independently of the set of parameters chosen (see Tab.~\ref{tab:xi}).
\begin{table}[ tp ]%
\begin{tabular}{c|c|c|c|c|c|c|c}
\hline %
Set & $p_{max}^{(s)}\ [\textrm{MeV}]$ &  $p_{max}^{(d)}\ [\textrm{MeV}]$ & $\gamma_{13}\ [\textrm{MeV}^{-3}]$ & $\gamma_{14}\ [\textrm{MeV}^{-3}]$ & $\gamma_{33}\ [\textrm{MeV}^{-5}]$ & $\gamma_{44}\ [\textrm{MeV}^{-5}]$ & $\gamma_{34}\ [\textrm{MeV}^{-5}]$\\\toprule %
1) & $1797.960$ & $868.265$ & $-0.875\cdot 10^{-7}$ & $1.169\cdot 10^{-7}$ & $-0.030\cdot 10^{-11}$ & $-0.055\cdot 10^{-11}$ & $0.003\cdot 10^{-11}$\\
2) & $1427.119$ & $865.693$ & $3.938\cdot 10^{-7}$ & $-5.028\cdot 10^{-7}$ & $-0.748\cdot 10^{-11}$ & $-1.345\cdot 10^{-11}$ & $0.966\cdot 10^{-11}$\\
3) & $1324.125$ & $904.062$ & $-0.695\cdot 10^{-7}$ & $0.840\cdot 10^{-7}$ & $-0.025\cdot 10^{-11}$ & $-0.033\cdot 10^{-11}$ & $-.002\cdot 10^{-11}$\\
4) & $1438.782$ & $897.246$ & $2.799\cdot 10^{-7}$ & $-3.502\cdot 10^{-7}$ & $-0.037\cdot 10^{-11}$ & $0.048\cdot 10^{-11}$ & $-0.028\cdot 10^{-11}$\\
5) & $1747.956$ & $911.004$ & $2.248\cdot 10^{-7}$ & $-2.873\cdot 10^{-7}$ & $0.126\cdot 10^{-11}$ & $0.162\cdot 10^{-11}$ & $-0.174\cdot 10^{-11}$\\
\hline
\end{tabular}
\caption{Values of the parameters of the theory resulting from a sample of five best fits to the scattering data for the $\bar{K}N$ and $\pi\Sigma$ amplitudes.}
\label{tab:fit}\centering %
\end{table}

At this point, we apply the sum rule of Eq.~\eqref{eq:sumrule2} to the present case. 
We first need to extrapolate the amplitudes to the complex plane and to look for the complex pole $\sqrt{s_0}$  in the second Riemann sheet. This is done by changing $G_i^{(s)}$ and $G_i^{(d)}$ to $G_i^{II (s)}$ and $G_i^{II (d)}$ in Eqs. \eqref{eq:loop_swave} and \eqref{eq:loop_dwave}, in the channels which are open. The functions $G_i^{II (s)}$ and $G_i^{II (d)}$ are the analytic continuations of the loop functions in the second Riemann sheet and are defined as
\begin{equation}
\begin{split}
&G_i^{II(s)}(\sqrt{s})=G_i^{I(s)}+\frac{i}{2\pi}\frac{M_i}{\sqrt{s}}\,q_i\ ,\\
&G_i^{II(d)}(\sqrt{s})=G_i^{I(d)}+\frac{i}{2\pi}\frac{M_i}{\sqrt{s}}\,\frac{q_i^5}{q_i^4(m_{\lambda^*})}\
,\ \ \ \ \ Im(q_i)>0\ ,
\label{eq:secondr}
\end{split}
\end{equation}
where $G_i^{I(s)}$ and $G_i^{I(d)}$ are the loop functions in the first Riemann sheet given by Eqs. \eqref{eq:loop_swave} and \eqref{eq:loop_dwave}. 

The values of the poles that we get from the five sets
are listed in the first column of Tab.~\ref{tab:coup}. 


Now we can evaluate the couplings of the resonance to the different channels as the residues at the pole of the amplitudes,
\begin{equation}
g_i^2=\lim_{\sqrt{s}\rightarrow \sqrt{s_0}}(\sqrt{s}-\sqrt{s_0})T_{ii}^{II}\ ,
\label{eq:coupling}
\end{equation}
and apply the sum rule to evaluate the contribution of a single channel to the $\Lambda(1520)$:
\begin{equation}
\label{eq:xi}
X_i=-Re\left[g_{i}^2\left[\frac{dG_i^{II}(s)}{d\sqrt{s}}\right]_{\sqrt{s}=\sqrt{s_0}}\right]\ .
\end{equation}

The couplings that we find using Eq.~\eqref{eq:coupling} are shown in Tab. \ref{tab:coup}.
Note that there is an ambiguity in the sign of $g_{\bar K N}$ and $g_{\pi\Sigma}$ among the
different sets but the product $g_{\bar K N}g_{\pi\Sigma}$ has the same sign.
 This is because we fit the transition $\bar K N\to \pi\Sigma$ which determines the relative sign
but not the overall one referred to the $\pi \Sigma^*$ channel.

\begin{table}[ tp ]%
\begin{tabular}{c|c|c|c|c|c|}
\hline %
 Set & $\sqrt{s_0}\ [\textrm{MeV}]$ &$g_{\pi\Sigma^*}$ & $g_{K\Xi^*}$ & $g_{\bar{K}N}$ & $g_{\pi\Sigma}$ \\\toprule %
$1)$ & $1518.7-i6.4$ & $0.70-i0.01$ & $-0.40+i0.05$ & $0.54 -i0.06$ & $0.43 -i0.05$\\
$2)$ & $1519.1-i6.7$ & $0.78-i0.07$ & $-0.35+i0.07$ & $-0.56+i0.05$ & $-0.45+i0.03$\\
$3)$ & $1518.3-i6.5$ & $0.73+i0.01$ & $-0.31+i0.03$ & $0.53 -i0.06$ & $0.44 -i0.06$\\
$4)$ & $1519.9-i6.5$ & $0.74+i0.00$ & $-0.34+i0.04$ & $0.53 -i0.07$ & $0.44 -i0.04$\\
$5)$ & $1518.5-i6.4$ & $0.63+i0.02$ & $-0.35+i0.03$ & $-0.53+i0.07$ & $-0.43+i0.05$\\
\hline
\end{tabular}
\caption{Pole positions and values of the couplings 
of the $\Lambda(1520)$ to the four different channels of the model.}
\label{tab:coup}\centering %
\end{table}

From these values we can obtain the relevance of the different channels in the wave function of the $\Lambda(1520)$ resonance, 
using Eq.~\eqref{eq:xi}. The values of the different weights are shown in Tab. \ref{tab:xi}.
\begin{table}[ tp ]%
\begin{tabular}{|c|c|c|c|c|c|}
\hline %
 Set &  $X_{\pi\Sigma^*}$ & $X_{K\Xi^*}$ & $X_{\bar{K}N}$ 
 & $X_{\pi\Sigma}$ & $1-Z$\\\toprule %
$1)$ & $0.084$ & $0.002$ & $0.494$ & $0.214$& $0.79$\\
$2)$ & $0.089$ & $0.001$ & $0.526$ & $0.225$& $0.84$\\
$3)$ & $0.093$ & $0.001$ & $0.541$ & $0.239$& $0.99$ \\
$4)$ & $0.093$ & $0.001$ & $0.518$ & $0.237$& $0.87$ \\
$5)$ & $0.072$ & $0.002$ & $0.531$ & $0.237$& $0.84$\\
\hline
\end{tabular}
\caption{Values of the weights $X_i$ 
of the different channels in the wave function of the $\Lambda(1520)$
and the total $1-Z=\sum_iX_i$} 
\label{tab:xi}\centering %
\end{table}

We can now estimate the composite character of the $\Lambda(1520)$
 resonance since, according to Eq.~\eqref{eq:sumrule2}
\begin{equation}
\label{eq:sr}
\sum_iX_i=1-Z\ ,
\end{equation}
where $Z$ is a measure of the presence in the state of something different
 from the meson-baryon components considered, (genuine components). 
 We obtain for $1-Z$ the values shown in the last column of Tab.~\ref{tab:xi}. 
Taking the average of
 the last column we have $1-Z=0.87\pm 0.10$,
 which indicate an appreciable weight of meson-baryon character
in the resonance with less than 15\% weight for other genuine components.
It is worth noting that numerically, the value of $Z$ corresponds~\cite{hyodorep} to
\be
\label{eq:hyodo}
Z=-\sum_{ij}\left[g_iG_i^{II}(\sqrt{s})
\frac{\partial V_{ij}(\sqrt{s})}{\partial \sqrt{s}}G_j^{II}(\sqrt{s})g_j\right]_
{\sqrt{s}=\sqrt{s_0}}.
\ee
Therefore, the diversion of $\sum_i X_i$ from unity is due to the smooth
energy dependence of the s-wave elements of the potential (see
 Eq.~\eqref{eq:potential}). In ref.~\cite{aceti2} it was shown that in cases where there is an explicit CDD pole in the
potential that accounts for a genuine state, or when one channel has
been eliminated introducing an equivalent effective potential, which
has a specific energy dependence, Eq. \eqref{eq:hyodo} indeed accounts for the
probability, or weight, of the missing channels. Yet, it is not clear
that the small magnitude obtained from the smooth energy dependence of
the Weinberg Tomozawa interaction can be attributed to missing
channels. We prefer to think that this amount can be considered as an
uncertainty in the method to determine $Z$. In the present case we also
see that this amount is of the same order of magnitude as the
statistical uncertainties. Anyway, the fact that we get a good fit without needing to include a CDD pole is an 
indication of the low weight of genuine components in the building up of the 
$\Lambda(1520)$ resonance.


On the other hand, we can see in Tab.~\ref{tab:coup} that the coupling of the resonance to
the $\pi\Sigma^{*}$ channel is the largest one. Yet, in terms of weight
(probability of the state if it was a bound state) it represents only
about $10\%$. This small probability can be deceiving, because the
relevance of each channel is usually tied to the values of the wave
function at the origin, more than to the probability. This is why in
each particular process one has to find out the relevance of each
channel. For instance, in the radiative decay
$\Lambda(1520)\rightarrow\gamma\Lambda,\,\gamma\Sigma^0$ it was found
that the $\pi\Sigma$, $\pi\Sigma^*$ channels did not contribute to the
$\gamma\Lambda$ decay, but in the case of $\gamma\Sigma^0$ decay channel
the $\pi\Sigma^*$ and $\pi\Sigma$ ($d$-waves) component gave the largest
contribution to the decay width \cite{michael}.

We should stress that one must be careful asserting the relevance of
the channels from the weight obtained. Indeed, for the open channels,
$\pi \Sigma$, $\bar K N$, the value of $X$ corresponds to the integral of
the wave function squared, which goes as $e^{-i q r}/r $ for large
$r$. While the integral of the modulus squared of the wave function
diverges, this is not the case for the wave function squared where the
oscillations of the $e^{-2i q r}$ factor lead to large cancellations
at large $r$. Yet, it is clear that for the open channels one is
getting contributions to $X$ from larger values of $r$ than in the bound
channels, $\pi \Sigma^*$, $K \Xi^*$. Yet, the wave function at large
values of $r$ will not have relevance in most processes involving short
distances.  In this sense, the couplings in the normalization that we
have, or the relative values of $X$ in the $s-$waves, or $d-$waves, channels
are the magnitudes that more fairly indicate the relevance of the
different channels, but ultimately it is the specific dynamics of a
given process that will determine the relevance of the channels, as
seen in \cite{michael}.


\section{Summary and conclusions}

We have applied the compositeness condition for resonances in higher partial waves to the case of
the $\Lambda(1520)$ resonance. The aim was to quantify the weight of the meson-baryon components
(s-waves $\pi\Sigma^*$, $K\Xi^*$ and d-waves $\bar{K}N$, $\pi\Sigma$) into the $\Lambda(1520)$ wave
function.
The meson-baryon scattering amplitudes are obtained implementing the techniques of the chiral
unitary approach where some unknown parameters (five  d-wave coefficients and two cutoffs)
 are fitted to $\bar{K}N$ and $\pi\Sigma$ experimental
scattering data.

The momentum dependence coming from the d-wave channels are incorporated into the loop function
leaving a smooth energy dependent potential for which the techniques developed 
in ref.~\cite{aceti} can be applied. From the knowledge of the 
loop functions and the couplings of the scattering
amplitudes to the different channels, obtained from the residues of the amplitudes at the pole
positions, the addends in the sum rule of Eq.~\eqref{eq:sumrule}, and the total sum rule itself,
can be evaluated which are a measure of the weight of the different channels into the  $\Lambda(1520)$
wave function. 

While the largest coupling obtained is to $\pi\Sigma^*$ (see Tab.~\ref{tab:coup}), the largest
weight $X_i$ is to $\bar K N$ (see Tab.~\ref{tab:xi}). This is not contradictory since they represent different concepts.
The coupling (actually the product of the coupling times the loop function, $g_iG_i$)
accounts for 
the wave function at the origin \cite{aceti} for $s-$waves while, as already explained, $X_i=-g_i\frac{\partial
G_i}{\partial E}$ is a measure of the probability to find that channel.

We also explained that the large weight obtained for the open
channels was a consequence of the contribution to the integral of the
wave function squared from larger values of $r$ than for the bound
channels, and not a measure of the contribution of the channel in
different processes, most of which are sensible to short distances.
The values of the couplings and the specific dynamics of those
processes are what finally determine the relevance of each of the
channels.



\section*{Acknowledgments}
This work is partly supported by the Spanish Ministerio de Economia y
Competitividad and European FEDER funds under Contract No.
FIS2011-28853-C02-01 and the Generalitat Valenciana in the program
Prometeo, 2009/090. We acknowledge the support of the European
Community-Research Infrastructure Integrating Activity Study of
Strongly Interacting Matter (Hadron Physics 3, Grant No. 283286) under
the Seventh Framework Programme of the European Union.

\end{document}